\documentclass[preprint,aps,tightenlines,floatfix,showpacs]{revtex4}

\usepackage{graphicx}
\usepackage{bm}
\usepackage{amsmath}
\usepackage{amssymb}

\begin{document}
\title{Nucleon-nucleus scattering as a test of shell structure of some 
light mass exotic nuclei}

\author{S. Karataglidis}
\email{S.Karataglidis@ru.ac.za}
\affiliation{Department of Physics and Electronics, Rhodes University,
P.O. Box 94, Grahamstown, 6140, South Africa.}

\author{Y. J. Kim}
\email{yjkim@cheju.ac.kr}
\affiliation{Department of Physics, Cheju National University,
Jeju 690-756, Republic of Korea.}

\author{K. Amos}
\email{amos@physics.unimelb.edu.au}
\affiliation{School of Physics, University of Melbourne,
Victoria 3010, Australia}

\date{\today}

\begin{abstract}
Shell  model  wave  functions  have  been  used  to  form  microscopic
$g$-folding optical potentials with which elastic scattering data from
$^8$He,  $^{10,11}$C, and $^{18,20,22}$O  scattering on  hydrogen have
been analyzed. Those potentials, the effective two-nucleon interaction
used in their  formation, and the shell model  details, then have been
used  in  distorted wave  approximation  calculations of  differential
cross sections  from inelastic scattering to the  first excited states
of five of those radioactive ions.
\end{abstract}


\maketitle


\section{Introduction}
Microscopic descriptions of exotic  nuclei are becoming more important
in  the  light  of  recent  experiments involving  the  scattering  of
radioactive ion beams (RIBs) off hydrogen. In inverse kinematics, such
equate  to scattering  of  protons  from the  ions,  and with  current
methods of  analysis~\cite{Am00} of  such scattering, a  most complete
map  of  their  matter  densities   can  be  made.  Such  well  tested
descriptions of exotic nuclei will be of relevance in analyses of data
to  be  taken  with  proposed  electron-ion collider  being  built  at
GSI~\cite{Si05},   as   well   as    from   the   SCRIT   project   at
RIKEN~\cite{Su05}. Descriptions of the charge densities, especially of
nuclei with neutron halos, will require detailed microscopic models to
account for the structure of the core.

Traditionally  proton scattering  has been  one of,  if not,  the best
means by  which the  matter densities of  the nucleus may  be studied.
Microscopic models now exist that  can predict results of both elastic
and inelastic scattering  reactions. When good, detailed specification
of the  nucleon structure  of the nucleus  is used,  those predictions
usually agree very well with observation, both in shape and magnitude.
To  facilitate such  analyses  of  data, one  first  must specify  the
nucleon-nucleus  ($NA$)  interaction.  To  do  so  requires  two  main
ingredients:  a)  an   effective  nucleon-nucleon  ($NN$)  interaction
in-medium,  allowing for  the  mean field  as  well as  Pauli-blocking
effects; and b) a credible model  of structure for the nucleus that is
nucleon-based. When the effective  $NN$ interaction is folded with the
one-body density matrix elements (OBDME) of the target ground state, a
microscopic $NA$ interaction results. Such interactions have been used
successfully  in  studies of  the  structures  of  stable nuclei  (see
\cite{Am00} for  a complete  review of those  studies), as well  as of
exotic nuclei~\cite{Ka00,La01}.

Herein,  we  consider the  scattering  of  the  exotic nuclei  $^8$He,
$^{10,11}$C,  and $^{18,20,22}$O  from hydrogen,  for which  there are
recent  data. The  data  for  the elastic  scattering  of $^8$He  from
hydrogen at $15.7A$~MeV  \cite{Sk05} was analyzed in terms  of the JLM
model using  $G$ matrix shell  model wave functions of  Navr\'atil and
Barrett  \cite{Na98}, as  well as  a coupled-channels  model involving
coupling to the  $^8$He($p,d$)$^7$He channel. Their analyses indicated
a  strong coupling  to the  ($p,d$)  channel. This  is problematic  as
$^7$He is  unbound. Any  possible recoupling to  reform $^8$He  in its
ground state is highly unlikely.  Indeed, their analysis based on this
coupling required  a scale strength of  the imaginary part  of the JLM
potential  of 0.2, which  the authors  attributed to  compound nucleus
effects. For $^8$He, the energy  (15.7 MeV) lies well in its continuum
and  compound nucleus contribution  to the  elastic scattering  is not
likely to be that large. So  such a sizable reduction in the imaginary
part of the JLM optical potential seems to be unrealistic.

The JLM interaction was also used in the analyses of the scattering of
the C~\cite{Jo05}  and O isotopes~\cite{Be06,Kh00}  from hydrogen. For
the analyses of the data from the C isotopes, and from $^{18,20}$O the
JLM potential was again used  with the adjustment of parameters to fit
the data. In the case of $^{22}$O \cite{Be06}, two analyses were made:
a)  one using  a folding  model producing  a real  microscopic optical
potential, but for which the imaginary part was obtained from a global
phenomenological parameterization; and b)  one using the JLM potential
for which  the neutron transition  densities were adjusted to  fit the
data.  As  a  consequence,  those  analyses  may  not  necessarily  be
sensitive  tests of  the structures  of  the targets  as any  possible
problems may  be masked by  a judicious use of  normalisation factors.
Thus we have  reanalyzed those data seeking a  better understanding of
the structures of the exotic ions. To do so we have used the Melbourne
$g$-folding   model   of   scattering~\cite{Am00}  together   with   a
microscopic distorted  wave approximation (DWA)  to describe inelastic
processes.  The shell  model has  been  used to  specify the  putative
structures of the ions.


\section{Structure and reaction theory}

In this section  we give details of the structure  of the exotic (RIB)
nuclei and of the theories used to evaluate elastic and inelastic data
from the scattering of those RIB nuclei from hydrogen.

\subsection{Many nucleon structures of $^8$He, $^{10,11}$C, and 
$^{18,20,22}$O} 

$^8$He  lies at  the neutron  drip  line. It  is weakly  bound with  a
two-neutron separation energy of  2.137~MeV~\cite{Aj88}. It is also an
example  of  a Borromean  nucleus  as $^7$He  is  unbound,  as is  any
two-body subsystem  of $^8$He. We have  used a no-core  shell model to
define  its  ground  state.  That  has been  obtained  in  a  complete
$(0+2+4)\hbar\omega$ model  space using the  $G$-matrix interaction of
Zheng \textit{et al.}~\cite{Zh95}. That ground state specification has
been   used   previously   in   analyses   of   scattering   data   at
$68A$~MeV~\cite{Ka00}.

\begin{figure}[ht]
\scalebox{0.45}{\includegraphics*{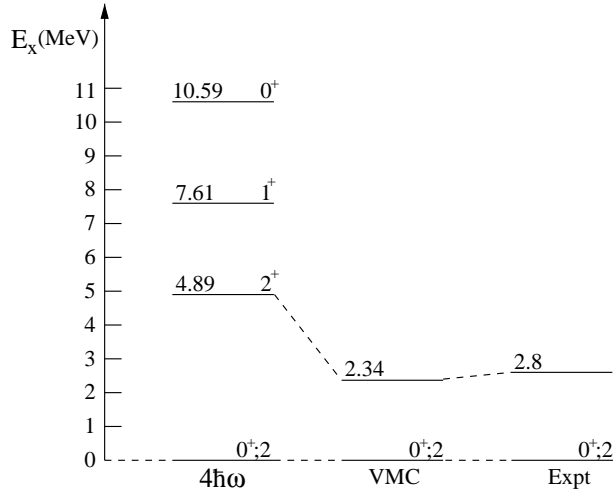}}
\caption[]{The spectrum of $^8$He. The result of the 
$(0+2+4)\hbar\omega$ shell model calculation is compared with that of 
the VMC calculation~\cite{Wi98}.}
\label{he8spec}
\end{figure}

Spectra for $^8$He are given in Fig.~\ref{he8spec}.
The  $(0+2+4)\hbar\omega$ shell  model results  are compared  not only
with  known data~\cite{Pu97}  but  also with  values  determined by  a
variational Monte  Carlo (VMC) calculation~\cite{Pu97}.  There is very
little experimental  information on the spectrum of  $^8$He. The first
excited  state is listed  at $2.8  \pm 0.4$~MeV  and has  $J^{\pi};T =
(2^+);2$~\cite{Aj88}.    Other    states,    at    1.3,    2.6,    and
4.0~MeV~\cite{Aj88},  have been  suggested from  a heavy  ion transfer
experiment, but as yet no other data support their existence.

The results  of the  VMC calculation place  the $2^+_1$ state  in very
close agreement  with experiment. However, that  calculation places an
extra $1^+$  state in the spectrum.  Not only has that  state not been
observed  in experiment, but  also it  has not  been found  with other
calculations. The shell model results clearly need improving to reduce
the gap  energy. Such suggest that  an even larger  space is required.
However, binding  energies reflect  the asymptotic properties  of wave
functions and  other tests  are required to  probe the  credibility of
current wave functions through the nuclear medium. A first test is the
root  mean   square  (r.m.s.)  radii,  which  is   most  sensitive  to
characteristics of the  (outer) surface of a nucleus.  Using our shell
model gave an  r.m.s. value for ${}^8$He of 2.63 fm,  which is in good
agreement with  the value  of 2.6 fm  extracted from high  energy data
using a cluster model~\cite{Al98}.

\begin{figure}
\scalebox{0.45}{\includegraphics*{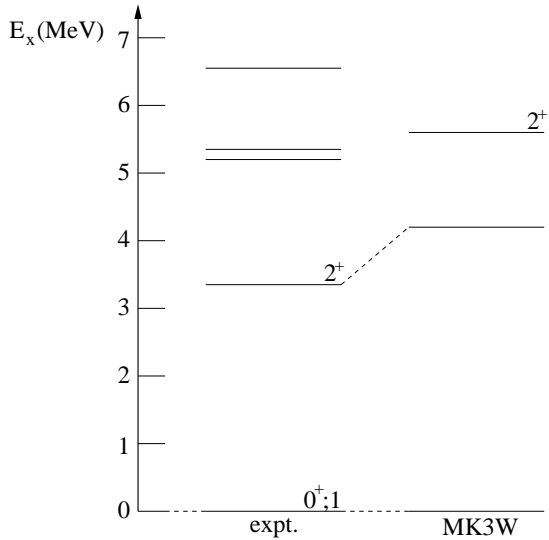}}
\caption{\label{c10spec}  Energy spectrum of  $^{10}$C. The  result of
the  $(0+2)\hbar\omega$  calculation  is   compared  to  the  data  of
Ref.~\cite{Ti04}.}
\end{figure}

Wave functions  of $^{10}$C and  $^{11}$C were obtained in  a complete
$(0+2)\hbar\omega$ model space using the MK3W interaction~\cite{Wa89}.
That  model reproduced  the $^{12}$C  spectrum to  20~MeV \cite{Ka95},
with the exception of the superdeformed $0^+$ state at 7.65~MeV. There
is  little  known  about the  spectrum  of  $^{10}$C  as is  shown  in
Fig.~\ref{c10spec}.
The  $2^+;1$ excited state  is at  3.354~MeV~\cite{Ti04} while  the SM
prediction puts it at 4.272~MeV.  The state at 6.58~MeV has been given
a tentative assignment  of $2^+$ for which the  corresponding state in
the SM  spectrum lies at 5.613~MeV.  There is no indication  in the SM
result  of  the  two  states  below  it.  They  have  not  been  given
assignments and may be intruder states.

\begin{figure}
\scalebox{0.45}{\includegraphics*{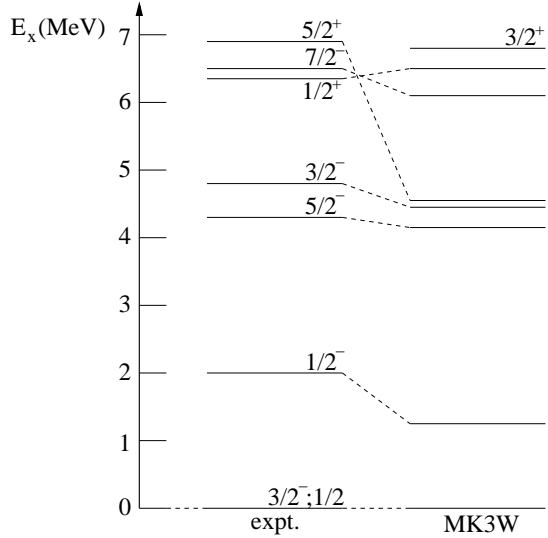}}
\caption{\label{c11spec} Energy spectrum of $^{11}$C. The 
results of the shell model calculations are compared to the data of
Ref.~\cite{Aj90}.}
\end{figure}

The spectrum of $^{11}$C is shown in Fig.~\ref{c11spec}.
Therein, the  results of $(0+2)\hbar\omega$  and $(1+3)\hbar\omega$ SM
calculations,  for negative and  positive parity  states respectively,
are compared  to the data~\cite{Aj90}. There is  better agreement with
the known spectrum in this case.

The  ground states for  $^{18,20,22}$O were  obtained from  a complete
$0\hbar\omega$ SM  calculation using the USD interaction  of Brown and
Wildenthal~\cite{Br88}. This  model of  structure suffices for  use in
describing  the elastic  scattering data  from these  nuclei.  For the
inelastic  scattering to  states  in $^{20}$O  and  $^{22}$O, we  also
obtained the transition densities in this model, noting that there may
be scaling of the cross sections due to core polarization corrections.

\subsection{The $g$-folding optical potential for elastic scattering}

A  detailed  study  of  the  $g$-folding optical  potential  has  been
published~\cite{Am00} and  so only salient features  are presented. In
coordinate space, that optical potential can be written as
\begin{equation}
U\left( \mathbf{r}, \mathbf{r}'; E \right) = \delta( \mathbf{r} -
\mathbf{r}' ) \sum_i n_i \int \varphi^{\ast}_i( \mathbf{r} )
g^D(\mathbf{r}, \mathbf{s}; E ) \varphi_i(\mathbf{s}) \, d\mathbf{s} +
\sum_i n_i \varphi^{\ast}_i( \mathbf{r} ) g^{E}( \mathbf{r},
\mathbf{r}'; E ) \varphi_i( \mathbf{r}' )
\end{equation}
To   evaluate  these  potentials   requires  specification   of  three
quantities.  They are the  single nucleon  bound state  wave functions
$\varphi_{i}(\mathbf{r})$,  the orbit  occupancies  $n_i$ which,  more
properly  are  the nuclear  state  OBDME,  and  the $NN$  $g$-matrices
$g^{D/E}(   \mathbf{r},\mathbf{r}';   E)$.   Those  $g$-matrices   are
appropriate combinations  of $NN$  interactions in the  nuclear medium
for  diverse  $NN$  angular  momentum  channels~\cite{Am00}.  For  the
latter, much  success has been had using  effective $NN$ interactions,
now commonly  designated as the  Melbourne force, which have  the form
$g^{ST}_{01}  \equiv  g^{ST}_{\text{eff}}\left(  r, E;  k_f(R)\right)$
where  $r =  \left|  \mathbf{r}_0  - \mathbf{r}_1  \right|$  and $R  =
\frac{1}{2} \left(  \mathbf{r}_0 + \mathbf{r}_1  \right)$. Therein the
Fermi momenta relate  to the local density in  the nucleus at distance
$R$ from  the center  when $\mathbf{r}_i$ are  the coordinates  of the
colliding projectile  and bound nucleons.  $\{ST\}$ are the  $NN$ spin
and isospin quantum numbers.

For use in the DWBA98  program, these effective $NN$ $g$-matrices are,
more specifically,
\begin{align}
g^{ST}_{\text{eff}}\left( \mathbf{r}, E; k_f \right) & = 
\sum^3_{i=1} \left[ \sum^4_{j=1} S^{(i)}_j \left( E; k_f \right)
\frac{e^{-\mu^{(i)}_j r}}{r} \right]_{\left[ S,T \right]} \Theta_i \nonumber \\
& = \sum_i g^{(i)ST}_{\text{eff}}\left( r, E; k_f \right) \Theta_i \; ,
\end{align}
where $\Theta_i$  are the characteristic operators  for central forces
($i =  1$), $\left\{ 1, (\mathbf{\sigma  \cdot \sigma}), (\mathbf{\tau
\cdot \tau}),(\mathbf{\sigma \cdot \sigma \tau \cdot \tau}) \right\}$,
for the  tensor force ($i  = 2$), $\left\{  \mathbf{S}_{12} \right\}$,
and for the  two-body spin-orbit force ($i =  3$), $\left\{ \mathbf{L}
\cdot  \mathbf{S} \right\}$. The  $S^{(i)}_j\left( E;k_f  \right)$ are
complex,  energy- and density-dependent  strengths. The  properties of
the $g$-matrices are such that  the ranges of the Yukawa form factors,
as assumed  in the above,  can be taken  as independent of  energy and
density~\cite{Am00}.

The strengths  (and ranges) in these effective  $NN$ interactions were
found  by mapping the  double Bessel  transforms of  them to  the $NN$
$g$-matrices  in infinite  nuclear matter  that are  solutions  of the
partial  wave  Bethe-Brueckner-Goldstone (BBG)  equations~\cite{Am00}.
With the BBG $g$-matrices denoted by $g^{JST}_{\text{BBG}; LL'}(q', q;
E)$, the mapping is
\begin{align}
g^{JST}_{\text{BBG}; LL'}(q', q; E)
& = \sum_i \left\langle \theta_i \right\rangle \int^{\infty}_0
r^{2+\lambda} j_{L'} \left( q'r \right) g^{(i)ST}_{\text{eff}}\left(
r, E; k_f \right) j_{L}\left( qr \right) \; dr \nonumber \\
& = \sum_{ij} \left\langle \theta_i \right\rangle S^{(i)}_j( \omega )
\int^{\infty}_0 r^{2+\lambda} j_{L'} \left( q'r \right) 
\frac{e^{-\mu^{(i)}_j r}}{r} j_{L}\left( qr \right) 
\; dr \nonumber \\
& = \sum_{ij} \left\langle \theta_i \right\rangle S^{(i)}_j( \omega )
\tau^{\alpha}\left( q', q; \mu^{(i)}_j \right) \; ,
\end{align}
where $\alpha: \{LL'JST\}$ and $\lambda = 0$ save for the tensor force
for which it is 2. In application, a singular-valued decomposition has
been used  to effect this  mapping and it  was found that  four Yukawa
functions for  the central  force and four  with other ranges  for the
spin-orbit and  tensor forces sufficed  to give close mapping  to both
on- and, a near range of,  off-shell values of the BBG $g$-matrices in
32 $NN$ $S, T$ channels and for energies to 300 MeV.

The other requirements come from  the assumed models of structure. The
many-nucleon aspects,  the OBDME, are the ground  state reduced matrix
elements of particle-hole operators
\begin{equation}
S_{jj0} = \left\langle
\Psi_{g.s.} \left\| \left[ a^{\dagger}_j \times {\tilde a}_j    
\right]^{0} \right\| \Psi_{g.s.} \right\rangle ,
\end{equation}
and which are  defined more generally in regard  to transitions in the
following subsection. If  there are no non-Hartree-Fock contributions,
these OBDME reduce  to the shell occupancies in  the ground state. The
single nucleon (bound state)  wave functions usually are chosen either
to be harmonic oscillators with oscillator length as used in the shell
model calculations  or from a Wood-Saxon potential.  The parameters of
that Wood-Saxon  potential are chosen to give  wave function solutions
that  meet some  criteria such  as the  r.m.s. radius  or  an electron
scattering form factor.

\subsection{The distorted wave approximation for inelastic scattering}

In the DWA, amplitudes for  inelastic scattering of nucleons through a
scattering angle $\theta$ and between states $J_i, M_i$ and $J_f, M_f$
in a nucleus, are given by
\begin{multline}
T^{M_fM_i\nu^\prime\nu}_{J_fJ_i}(\theta) = \\ 
\left\langle \chi^{(-)}_{\nu^\prime}({\bf k}_o0)\right|
\left\langle\Psi_{J_fM_f}(1 \cdots A) \right|
A \sum_{ST} \, g_{\text{eff}}^{ST}(r_{01}, E; k_f)\,P_S P_T
\mathcal{A}_{01} \left\{ \left| \chi^{(+)}_\nu ({\bf
k}_i0) \right\rangle \left| \Psi_{J_iM_i}
 (1\cdots A) \right\rangle \right\} ,
\end{multline}
where $\nu, \nu^\prime$ are the spin quantum numbers of the nucleon in
the   continuum,   $\chi^{(\pm)}$  are   the   distorted  waves,   and
$g_{\text{eff}}^{ST}(r_{01}, E; k_f)\, P_S  P_T $ is the spin-isospin
Melbourne    force.    The    operator   $\mathcal{A}_{01}$    effects
antisymmetrization of the two-nucleon product states.

Then, by using cofactor expansions of the nuclear states, i.e.
\begin{equation}
\left| \Psi_{JM} \right\rangle = \frac{1}{\sqrt{A}} \sum_{j,m}
\left| \varphi_{jm}(1) \right\rangle a_{jm} \left| \Psi_{JM}
\right\rangle ,
\end{equation}
the matrix elements become
\begin{multline}
T^{M_f M_i \nu \nu'}_{J_f J_i}( \theta ) = 
\sum_{j_1 j_2 S T } \left\langle J_f M_f \left| a^{\dagger}_{j_2 m_2}
a_{j_1 m_1} \right| J_i M_i \right\rangle \\
\times \left\langle \chi^{(-)}_{\nu'}( \mathbf{k}_o 0 ) \right|
\left\langle \varphi_{j_2 m_2}(1) \right| g^{ST}_{\text{eff}}\left( 
r_{01}, E; k_f \right) 
P_S P_T \mathcal{A}_{01} \left\{ \left| \chi_{\nu}^{(+)} \left(
\mathbf{k}_i 0 \right) \right\rangle \left| \varphi_{j_1 m_1}(1)
\right\rangle \right\} \; .
\end{multline}
The density matrix elements in the amplitudes reduce as
\begin{align}
\lefteqn{ \left\langle J_f M_f \left| a^{\dagger}_{j_2 m_2}
\times a_{j_1 m_1} \right| J_i M_i \right\rangle} \nonumber \\
& = \sum_{I(N)} (-1)^{j_1 - m_1} \left\langle j_1 \, m_1 \, j_2 \,
-m_2 | I \, N \right\rangle \left\langle J_f M_f \left| \left[ a^{\dagger}_{j_2}
\times \tilde{a}_{j_1}
\right]^{IN} \right| J_i M_i \right\rangle \nonumber \\
& = \sum_{I(N)} \frac{(-1)^{j_1 - m_1}}{\sqrt{2J_f + 1}}
\left\langle j_1 \, m_1 \, j_2 \, -m_2 | I \, N \right\rangle
\left\langle J_i \, M_i \, I \, N | J_f \, M_f \right\rangle S_{j_1 j_2 I}\; ,
\end{align}
where $S_{j_1j_2I}$ are the transition OBDME.   The DWA amplitudes are 
then (with $\xi = \left\{  j_1,j_2,m_1,m_2,I(N),S,T \right\}$)
\begin{multline}
T^{M_fM_i \nu' \nu}_{J_f J_i}(\theta) = 
\sum_{\xi} \frac{(-1)^{j_1 - m_1}}{\sqrt{2J_f + 1}} S_{j_1j_2I}
\left\langle j_1 \, m_1 \, j_2 \, -m_2 | I \, N \right\rangle
\left\langle J_i \, M_i \, I \, N | J_f \, M_f \right\rangle \\
\times \left\langle \chi^{(-)}_{\nu'} \left( \mathbf{k}_o 0 \right) \right|
\left\langle \varphi_{j_2 m_2}(1) \right| g^{ST}_{\text{eff}}\left(
r_{01}, E; k_f \right)  P_SP_T \mathcal{A}_{01} \left\{ \left| \chi^{(+)}_{\nu}\left(
\mathbf{k}_i 0 \right) \right\rangle \left| \varphi_{j_1 m_1}(1) \right\rangle
\right\} \, .
\end{multline}

In our calculations of these  DWA amplitudes, a) the $g$-folding model
has been  used to determine the  distorted waves in  both the incident
and emergent channels, b) the same effective $NN$ interaction has been
used as  the transition  operator, c) shell  models have been  used to
find the  transition OBDME, and  d) the single nucleon  wave functions
chosen  to form the  optical potentials  have also  been used  for the
single nucleon states.

\section{Results}

In  previous  papers~\cite{La01,St02},  the  cross sections  from  the
elastic and inelastic scattering (to the $2_1^+$ state) of $40.9A$ and
$24.5A$~MeV $^6$He ions from  hydrogen were analyzed using the methods
described above. Those results provided clear evidence that $^6$He had
an  extended neutron  matter  distribution consistent  with a  neutron
halo. Similar analyses of  elastic scattering data from the scattering
of $72A$ MeV $^8$He ions from hydrogen suggested that it had a neutron
skin  but  not  the   extended  distribution  of  a  halo~\cite{Ka00}.
Recently~\cite{Sk05}  data from the  scattering of  $15.7A$~MeV $^8$He
ions from  hydrogen has been reported,  and we consider  that now. The
energy  is the  smallest  at  which the  $g$-folding  method has  been
applied. In  most nuclei in the  excitation energy range  $\sim 10$ to
$\sim 30$~MeV  there are  giant resonances; the  coupling to  which is
known~\cite{Ge75} to  influence nucleon  scattering. When that  is so,
corrections  to the  $g$-folding  premise need  be  made. However,  an
important exception is the set  of Helium isotopes for which there are
no giant resonances. With  $^8$He also, the nucleon break-up threshold
lies at just a few MeV so  that 15~MeV is well in the continuum of the
nucleus.  For these  reasons,  it  should be  appropriate  to use  the
$g$-folding approach to analyze the $15.7A$~MeV data.

\subsection{The elastic scattering of $15.7A$ MeV $^8$He from hydrogen}

\begin{figure}[ht]
\scalebox{0.5}{\includegraphics{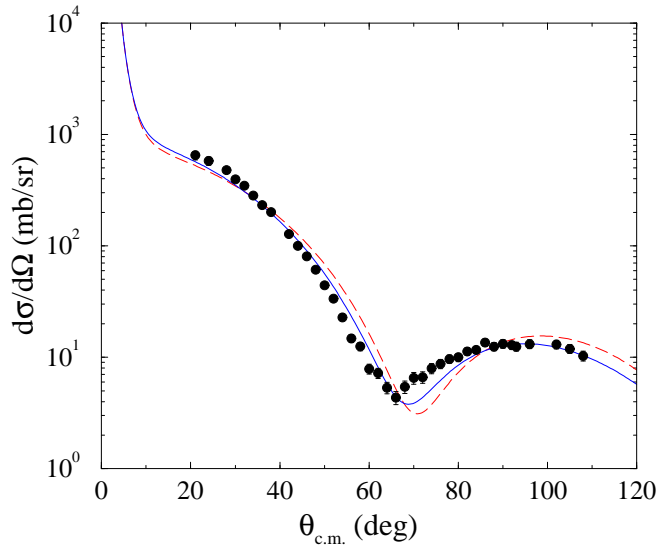}}
\caption{\label{Fig1} (Color  online) Differential cross  sections for
the elastic  scattering of $15.7A$~MeV $^8$He ions  from hydrogen. The
data~\cite{Sk05}  are  compared with  $g$-folding  model results.  The
curves are identified in the text.}
\end{figure}

Using the structure of the  ground state described previously and with
the  15  MeV  Melbourne  force,  two predictions  of  the  $15.7A$~MeV
$^8$He-$p$ scattering  have been made.  The results are  compared with
the data in Fig.~\ref{Fig1}.
The result  shown by  the dashed curve  was found by  using oscillator
wave functions for  the single nucleon states in  the folding process,
while  that shown  by  the  solid curve  was  found using  Woods-Saxon
functions generated  with the potential  used before~\cite{La01,St02}.
Clearly the predictions are in very good agreement with this data with
preference (as  was found  before) for the  Woods-Saxon single-nucleon
wave  function set.  The JLM  model results~\cite{Sk05}  do  better in
reproducing this data around $80^\circ$  but do not as well at forward
scattering angles.  However, the  JLM results required  some \textit{a
posteriori}  adjustment of  the real  and imaginary  strengths  of the
optical  potential  from  those  considered standard.  That,  and  the
strength  of the  measured $^8$He($p,d$)  cross section,  led  them to
believe that a second order  $(p,d)(d,p)$ process was important in the
description of elastic scattering. But, the measurement of the ($p,d$)
reaction was inclusive with no identification of the final state being
$^7$He. As $^7$He  is unbound against neutron emission  there are many
multi-particle  exit  channels  from  the  ($p,d$)  reaction  and  the
likelihood of  the conglomerate  reforming to a  $^8$He and  proton is
very small.  In fact the  plethora of propagating channels  (given the
energy and  momentum sharing  possible) is one  reason why  an optical
potential approach is appropriate.

\subsection{The scattering of $^{10,11}$C ions from hydrogen}

\begin{figure}[ht]
\scalebox{0.5}{\includegraphics{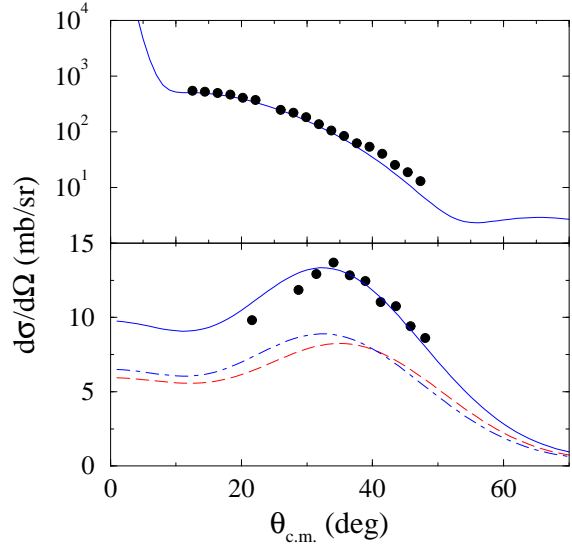}}
\caption{\label{Fig2} (Color  online) Differential cross  sections for
the  elastic  (top) and  inelastic  scattering  to  the $2_1^+$  state
(bottom) of $45.3A$ MeV $^{10}$C ions from hydrogen.}
\end{figure}

\begin{figure}[ht]
\scalebox{0.5}{\includegraphics{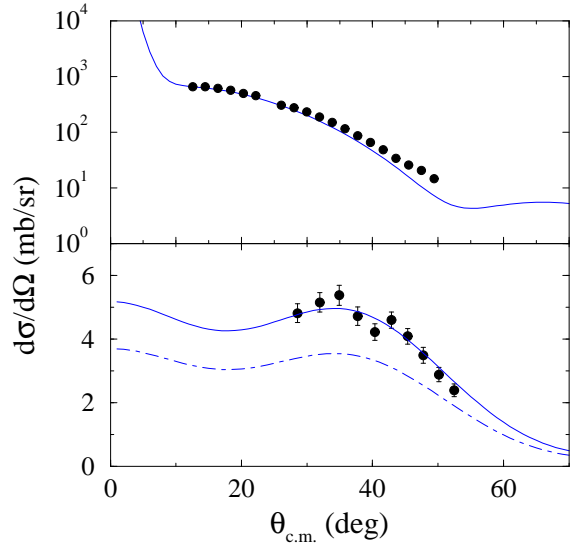}}
\caption{\label{Fig3} (Color  online) Differential cross  sections for
the  elastic (top)  and  inelastic scattering  to the  $\frac{5}{2}^-$
state (bottom) of $40.6A$ MeV $^{11}$C ions from hydrogen.}
\end{figure}

Another  GANIL experiment  \cite{Jo05}  determined differential  cross
sections  for  the elastic  and  inelastic  scattering of  $45.3A$~MeV
$^{10}$C  ions and of  $40.6A$~MeV $^{11}$C  ions from  hydrogen. That
data are compared with predictions in Figs.~\ref{Fig2} and \ref{Fig3}.
Previously, data taken at  $40.9A$ MeV~\cite{La01,St02} of $^6$He ions
elastically and  inelastically scattered from  hydrogen were predicted
well with the $g$-folding method.  The current results, with which the
new data  are compared in these  two figures, were found  by using the
structure as described  earlier. Also we have used  the same Melbourne
force   that    was   used   in    the   analyses   of    the   $^6$He
data~\cite{La01,St02}  and  either   oscillator  or  Woods-Saxon  wave
functions for the bound nucleons.
Note that in these diagrams the inelastic scattering data and  results 
are plotted on a linear scale to emphasize any shortfalls.

In Fig.~\ref{Fig2}, cross sections  for the scattering of $^{10}$C are
shown. Clearly the elastic  scattering prediction agrees with the data
almost  as  well  as  the  phenomenological JLM  analyses  of  Jouanne
\textit{et  al.}~\cite{Jo05}. With  the inelastic  scattering  (to the
$2^+$; 3.36  MeV state),  the results found  using the OBDME  from the
shell model calculations described before and with oscillator (dashed)
and Woods-Saxon  (dot-dashed) functions for  the single-nucleon states
required in the DWA calculations  lie below the data values. A scaling
of 1.5  on the Woods-Saxon  result gives the  solid curve which  is in
very good agreement with the  data. Scaling was also required with the
phenomenological  model analyses~\cite{Jo05}.  In  our case,  however,
given  the  propriety (admittedly  mostly  justified through  numerous
uses) of
\begin{enumerate}
\item 
the  model, and  in  particular  with it,  specific  treatment of  the
exchange amplitudes;
\item
the transition operator (Melbourne force) in which account is taken of
medium effects on the $NN$ interaction; and
\item
the  single-nucleon wave  functions which  link well  to  ground state
properties;
\end{enumerate}
the  shortfall  we  attribute  to  a limitation  of  the  many-nucleon
structure for the ion itself.  It has been long known~\cite{Am00} that
core polarization  corrections, be they made  phenomenologically or by
appropriate increase in the space  in which the structure is assessed,
influence   inelastic  scattering   evaluations  strongly;   and  most
obviously when collective enhancing excitations are involved. That the
amount in  the case of $^{10}$C  is but a 50\%  enlargement attests to
the reasonable first guess that is given by the structure calculations
made  using  the  MK3W   potentials.  The  enlargement  equates  to  a
polarization charge, $\Delta e \approx 0.11 e$.

In Fig.~\ref{Fig3},  the $40.6A$ MeV  $^{11}$C results are  shown. The
elastic scattering  data are again  well predicted by  the $g$-folding
method.   Likewise   the  cross   section   for   excitation  of   the
$\frac{5}{2}^-$; 4.32  MeV state  as predicted using  Woods-Saxon wave
functions (dot-dashed  curve) reproduces  the observed data  very well
when a small upward scale (of 1.4) is made.

Thus the  MK3W structure  adopted for these  states in  $^{10,11}$C is
quite good. Use of that  structure in the scattering calculations gave
cross-section shapes very like that observed in the data and requiring
but a small amount of core polarization to reproduce the measured data
from the  inelastic ($E2$ dominant)  transitions. We surmise  that the
first  complete basis space  enlargement possible  to the  shell model
scheme we  have used to describe  these nuclei will give  much, if not
all, of such enhancement.

\subsection{The $^{18,20,22}$O isotopes}

\begin{figure}[ht]
\scalebox{0.5}{\includegraphics{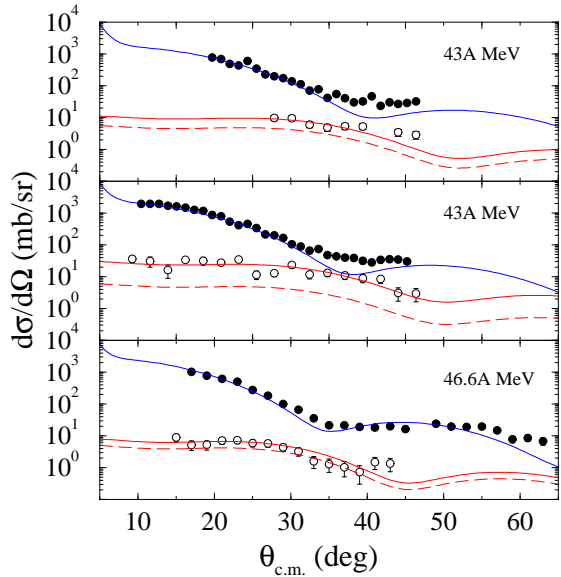}}
\caption{\label{Fig4} (Color  online) Differential cross  sections for
the elastic and inelastic scattering  (to the $2_1^+$ states) with the
isotopes  of oxygen,  top  ($^{18}$O), middle  ($^{20}$O), and  bottom
($^{22}$O). The energies are as indicated.}
\end{figure}

Recently the  cross sections from $43A$~MeV $^{18,20}$O  ions and from
$46.6A$~MeV $^{22}$O  ions scattering from hydrogen  targets have been
measured~\cite{Kh00,Be06}. Cross sections for both the elastic and the
inelastic excitation  of $2^+_1$ states  of those ions  were obtained.
Those  data  are  shown in  the  top,  middle,  and bottom  panels  of
Fig.~\ref{Fig4}.
Those  data are  compared with  our predictions  ($g$-folding  for the
elastic  and  DWA  for  the  inelastic  transitions)  built  with  the
$0\hbar\omega$ shell model wave functions and the Melbourne force. The
elastic  scattering cross-section  predictions depicted  by  the solid
lines  agree  well  with  the  data (filled  circles)  save  that  the
$^{18,20}$O results  have an over-pronounced  minimum near $40^\circ$.
The  inelastic scattering  cross sections  given by  our  approach are
those depicted by the long dashed curves and they all are smaller than
any of the data sets (opaque circles). The solid curves that match the
inelastic data well are those calculated values increase by factors of
2 ($^{18}$O), 5 ($^{20}$O), and 1.6 ($^{22}$O). These ratios are quite
similar  to those of  the squares  of the  deformations required  in a
previous analysis  of this data~\cite{Gu06}. That  analysis also found
very good  fits to all  of the elastic  scattering data, but to  do so
involved using  many parameters including  separate renormalization of
the real and  imaginary components of the optical  potentials, and for
each nucleus independently.

With our results  for the inelastic scattering cross  sections one can
envisage different  degrees of  core polarization being  required with
the  base  valence  models  having   2,  4,  and  6  (full  sub-shell)
$d_{\frac{5}{2}}$-orbit  occupancies  as  the  dominant  term  in  the
description of the ground  states of $^{18,20,22}$O respectively. That
$^{20}$O requires largest scaling is reflective of that nucleus having
considerably more  configuration mixing than  the other isotopes  in a
good description of its states.

\section{Conclusions}

The  microscopic $g$-folding  approach to  analysis of  data  from the
elastic scattering of radioactive ion beams from hydrogen targets when
the beam energy equates to an excitation of those ions sufficiently in
their continua,  is most  appropriate. When essentials  are satisfied,
that approach  can, and does,  give predictions of  elastic scattering
that  one  may have  confidence  in  being  a good  representation  of
observation. It is  essential, though, to use a  realistic energy- and
medium-dependent effective $NN$ interaction in the folding. Also it is
essential  that as good  a prescription  as possible  be made  for the
distributions of nucleons in the  ground state of the ion. Finally one
must be faithful to the Pauli principle, not only in the specification
of the effective $NN$ interaction,  but also in the calculation of the
scattering allowing  for the knock-out process. Doing  so introduces a
specific non-local  term in the  optical potential. Our  experience is
that such  non-locality should be  treated as exactly as  possible and
not localized  to simplify the problem  to solution of a  set of local
Schr\"odinger equations.

We  have also shown  that the  DWA suffices  to analyze  the inelastic
scattering  cross sections.  There is  a  proviso that  the energy  is
sufficient that  coupled-channel effects due to  discrete and/or giant
resonance states  of the ion are  not important. When such  is not the
case, other methods of analysis, such as with a multichannel algebraic
scattering  theory~\cite{Ca06}, are  more relevant.  Even then  it has
been   noted    that   due   care   of   the    Pauli   principle   is
essential~\cite{Am05}. But  for data  at the energies  considered, the
DWA worked  very well.  With all details  preset, the DWA  results are
also predictions  which, in the cases considered,  well reproduced the
shapes of the data. All of those calculated results had to be enhanced
to meet the magnitudes of that data but those scalings were consistent
with  the core  polarization  one  must expect  for  shell model  wave
functions defined in the small bases we have used.

\bibliography{HeCO-paper}

\end{document}